# Power System Voltage Stability Boundary: Computational Results and Applications

Zhenyao Li, Yifan Yao, Deqiang Gan, *Senior Member*, *IEEE*

*Abstract*—The objective of this paper is to report some computational results for the theory of DAE stability boundary, with the aim of advancing applications in power system voltage stability studies. Firstly, a new regularization transformation for standard differential-algebraic equations (DAEs) is proposed. Then the existence of anchor points on voltage stability boundary is examined, and an optimization method for computing the controlling pseudo-saddle is suggested. Subsequently, a local representation of the stable manifold of the pseudo-saddle on the stability boundary is presented, and a voltage stability margin expression is obtained. Finally, the proposed results are verified using several examples, demonstrating the accuracy and effectiveness of the suggested methods.

*Index Terms*—short-term voltage stability, regularization transformation, impasse surface, stability region, stability boundary.[1]

## I. Introduction

The dynamics of a power system is described by a set of differential algebraic equations (DAEs) [1]. It is a common practice to determine the voltage stability of a power system by observing system trajectories after the occurrence of a disturbance. Particularly, voltage stability is closely related to whether the trajectories reach the singular surface, i.e., the Jacobian matrix of the algebraic equations with respect to the algebraic variables being singular [2].

When the constant impedance load model is used to represent loads, the power system is usually free of singularity problem. Whereas, with the adoption of the nonlinear load model, power systems begin to develop singular surfaces which will expand as the proportion of the nonlinear load increases [3]. The behavior of grid-following (GFL) converters is similar to that of the constant power loads. In contrast, the dynamic behavior of the grid-forming (GFM) converters is similar to that of a controllable voltage source [4], which can in general improve voltage stability.

The research interest in the notion of the singular surface focuses on the conditions under which the system trajectory hits the singular surface [5],[6]. This is also named as the solvability conditions of the algebraic equations of the power system models in some literature. An energy function transient stability method was developed in [7] to find the point of intersection with the singular surface. The impact of load variation on the system's singularity condition was revealed in [8]. An idea of quasi-dynamics was proposed in [9] to analyze the effect of various loads on the voltage impasse region.

The relevance of singular surface in voltage stability studies can be explained using the notion of stability boundary of nonlinear systems. Chiang and his co-workers established the theory of the stability region of ordinary differential equations (ODEs) in the 1980s [10],[11]. Later in the early 1990s, Venkatasubramanian and his colleagues [12],[13] developed a theory for stability boundaries of DAEs based on the so-called regularization transformation. The main result of the theory, geometric in nature, states that the stable boundary of a DAE system is roughly composed of the stable manifolds of certain anchor points (such as the unstable equilibrium points, pseudo-equilibrium points and semi-singular points) and parts of singular surface. This result can also be established using a singular perturbation argument [14]. It precisely explains why the singular surface is of significance in voltage stability studies [15]. Besides, the theory also extends naturally to systems with inequality state constraints [16].

From an application perspective, a large number of literature exist about the computation of ODE stability boundary, the readers are referred to (say) [11],[17]-[19] for various developments on the subject. In contrast, there exist very few results dealing with the computation of DAE stability boundary. The behaviors of instability of power systems interconnected with renewable resources were investigated in [20],[21], while our recent work proposed a singularity based voltage stability indicator and revealed the graph-theoretic properties of singularity conditions [15].

The main thrust of this work is to develop some computational results for DAE stability boundary theory, with applications to power system voltage study. Of particular interest is the computation of pseudo-saddles and their stable manifolds that have a detrimental impact on voltage stability behavior. A minimum modulus eigenvalue-based regularization transformation is introduced to secure desirable numerical stability. It is demonstrated that the new transformation is topologically equivalent to the standard

[1] This research is support by State Grid Corporation under Headquarter Fundamental Research Initiative (Stability region algebraic properties and transient voltage instability of hybrid power systems, 5100-202499307A-1-3-ZB).

The authors are with the College of Electrical Engineering, Zhejiang University, Hangzhou, China. (Emails: {12110085, yyf2000, dgan}@zju.edu.cn)

determinant-based transformation. The existence of anchor points on the stability boundary of the transformed system is examined, and a method for computing the controlling pseudo-saddle is proposed. A local representation of the stable manifold of the pseudo-saddle is derived, this allows us to establish a new voltage stability indicator which together with singularity indicators provides a more complete understanding into the complex behavior of voltage dynamics. Finally, the proposed results are verified using several examples, demonstrating the accuracy and effectiveness of the suggested methods.

## II. THE CONSTRAINED POWER SYSTEM MODEL

This section gives the constrained power system model. The constrained power system model is constituted by differential equations that describe the dynamics of the state variables and algebraic equations that describe transmission network constraints.

### A. Differential Equations of Synchronous Machines and GFM Converters

The differential equations mainly describe the dynamics of the synchronous machines and GFM converters.

The model for synchronous machines is as follows

$$f_1 : \begin{cases} \dot{\delta} = \omega_0 \omega \\ M\dot{\omega} = P_m - P_e - D\omega \\ T'_{d0}\dot{E}'_q = E_{fd} - E'_q - (x_d - x'_d)I_d \\ T_A \dot{E}_{fd} = K_A(V_{ref} - |V_G|) - E_{fd} \end{cases} \quad (1)$$

where $\delta$ denotes the rotor angle, $\omega$ and $\omega_0$ denote the frequency of machines and the synchronous frequency, $M$ denotes the inertia constant, $P_m$ and $P_e$ denote the mechanical power and the electromagnetic power, $D$ denotes the damping coefficient, $T'_{d0}$ denotes the open-circuit time constant, $E'_q$ denotes the q-axis component of the voltage behind transient reactance, $E_{fd}$ denotes the excitation voltage, $x_d$ and $x'_d$ denote the d-axis synchronous reactance and the d-axis transient reactance, $I_d$ denotes the d-axis current, $T_A$ denotes the time constant, $K_A$ denotes the excitation gain, $V_{ref}$ denotes the reference voltage, $V_G$ denotes the terminal voltage of a synchronous machine, "./" denotes element-wise division operations (following the MATLAB notation).

Virtual synchronous machine (VSM) is a major control strategy for GFM converters at present. This paper presents a case study of VSM as a representative example, and other control strategies can be validated through the same methodology.

The model of GFM converters for the VSM strategy is as follows (see Fig. 16 in Section VI.D)

$$f_2 : \begin{cases} \dot{\delta} = \omega_0 \omega \\ M\dot{\omega} = P_{ref} - P_{GFM} - D\omega \\ K_i \dot{E}_C = E_{Cfd} + K_q(Q_{ref} - Q_e) \\ T_u \dot{E}_{Cfd} = V_{ref} - |V_G| - E_{Cfd} \end{cases} \quad (2)$$

where, $P_{ref}$ and $P_{GFM}$ denote the reference power and the active power of the GFM converter, $x_l$ denotes the outlet impedance of the converter, $E_C$ denotes the internal voltage of the converter, $E_{Cfd}$ denotes the virtual excitation voltage of the converter, $K_i$ and $T_u$ denote the time constant, $Q_{ref}$ and $Q_e$ denote the reference reactive power and output reactive power of the converter, $K_q$ denotes the Q-V droop gain.

### B. Algebraic Equations of The Transmission Network Constraints

All currents injected into the grid need to satisfy the following transmission network constraints

$$YV = I \quad (3)$$

where $Y$ denotes the reduced network admittance matrix after eliminating the buses without current injected in, $V$ denotes the bus voltages and $I$ denotes the currents injected into the buses. To be specific, the buses were divided into generator buses and load buses, and let $V_G$ and $V_L$ denote their corresponding voltages. Since the transient potentials of synchronous machines and the internal voltages of GFM converters exhibits similar dynamics characteristics, let $E$ denote both of them. GFL converters are regarded as constant power loads that inject current into the grid. Then it can be obtained that [22],[23]

$$\begin{bmatrix} Y_{GG} & Y_{GL} \\ Y_{LG} & Y_{LL} \end{bmatrix} \begin{bmatrix} V_G \\ V_L \end{bmatrix} = \begin{bmatrix} -j(E \circ e^{j\delta})./x'_d \\ (\rho \circ \bar{S}_L)./\bar{V}_L + Y_z V_L \end{bmatrix} \quad (4)$$

where, $Y_{GG}$, $Y_{GL}$, $Y_{LG}$ and $Y_{LL}$ denote the Kron-reduced admittance matrices, $\rho$ denotes the percentage of constant power loads in the loads, $S_L$ denotes the complex power of the loads, $Y_z$ denotes the constant impedance loads and it was calculated as

$$Y_z = \text{diag}[(I - \rho) \circ \bar{S}_L./|V_{L0}|^2] \quad (5)$$

where diag(·) denotes the function of a diagonal matrix consisting of elements in parentheses, $V_{L0}$ denotes the load bus voltages, the values of which are obtained from power flow solution, '∘' denotes the Hadamard product.

Notice that

$$Y = \begin{bmatrix} Y_{GG} & Y_{GL} \\ Y_{LG} & Y_{LL} \end{bmatrix}, V = \begin{bmatrix} V_G \\ V_L \end{bmatrix} \quad (6)$$

let

$$Y_{LE} = \begin{bmatrix} \text{diag}(jx'^{-1}_d) \\ 0 \end{bmatrix}, \bar{S}'_L = \begin{bmatrix} 0 \\ \rho \circ \bar{S}_L \end{bmatrix}, Y'_z = \begin{bmatrix} 0 & 0 \\ 0 & Y_z \end{bmatrix} \quad (7)$$

then it can be obtained that

$$jB_{LE}(E \circ e^{j\delta}) + (G + jB)V = \bar{S}'_L./\bar{V} \quad (8)$$

where

$$jB_{LE} = Y_{LE}, G + jB = Y - Y'_z \quad (9)$$

Separating the real and imaginary parts in (8), the algebraic equations of the transmission network constraints can be obtained as follows

$$g : \begin{cases} 0 = BV_x + GV_y + I_{Gx} - I_{Lx} \\ 0 = GV_x - BV_y + I_{Gy} - I_{Ly} \end{cases} \quad (10)$$

where $V_x$ and $V_y$ denote the components of $V$ on the x-y axis, and

$$I_{Gx} = B_{LE}(E \circ \cos \delta)$$
$$I_{Gy} = -B_{LE}(E \circ \sin \delta)$$
$$I_{Lx} = (PV_y - QV_x)./(V_x^2 + V_y^2) \quad (11)$$
$$I_{Ly} = (PV_x + QV_y)./(V_x^2 + V_y^2)$$

In the above equations, $P$ and $Q$ denote the active power and reactive power of the constant power loads. For the sake of simplicity, the parameter $\rho$ is omitted in subsequent text.

In summary, the constrained power system model is composed of (1), (2) and (10). From (10), it can be seen that when there is no constant power load in a power system, the algebraic equations must have explicit solutions and the power system can be modeled by ODEs. Conversely, when there are constant power loads, it is in general impossible to obtain an explicit solution of the algebraic equation. The next section will describe regularization transformations that help convert DAE models into equivalent ODE models.

### III. A New Regularization Transformation for General DAEs

This section first revisits the standard conception of the regularization transformation which is based on the determinant of the Jacobian matrix of network equations [13],[24], and then proposes a transformation based on the minimum modulus eigenvalue of the Jacobian matrix. This new transformation technique is applicable to general DAEs.

#### A. Regularization Transformation Based on Determinant

The power system model is a typical constrained nonlinear dynamical system as follows

$$(\Sigma) \begin{cases} \dot{x} = f(x, y) \\ 0 = g(x, y) \end{cases} \quad (12)$$

where $x \in R^n$ denotes the vector of the state variables, $y \in R^m$ denotes the vector of the algebraic variables, $f$ and $g$ are smooth functions.

The constrain set $\mathcal{L}$ of the system is defined as

$$\mathcal{L} = \{(x, y) \in R^{n+m} | g(x, y) = 0\} \quad (13)$$

When the loads are modeled as constant impedance loads, the matrix $D_y g$ (the derivative of $g$ with respect to $y$) is constant and invertible. Consequently, by the implicit function theorem, the algebraic constraint $y$ can be solved as $y = g^{-1}(x)$.

Nevertheless, the explicit expression of $g^{-1}(x)$ is typically unavailable when nonlinear components are present. In particular, when matrix $D_y g$ is non-invertible, the system reaches the singular surface $\mathcal{S}$ which is defined as

$$\mathcal{S} = \{(x, y) \in \mathcal{L} | \Delta(x, y) := \det D_y g = 0\} \quad (14)$$

where $\det(D_y g)$ denotes the determinant of $D_y g$.

One of the existing approaches to analyzing the stability region and stability boundary characteristics of a DAE system is to construct a well-defined transformed vector field in the singular surface, which is equivalent to the DAE system in the set $\mathcal{L} \backslash \mathcal{S}$. To proceed, the algebraic equation can be differentiated to yield

$$(D_x g)f + (D_y g)\dot{y} = 0 \quad (15)$$

where $D_x g$ denotes the derivative of $g$ with respect to $x$. Suppose $D_y g$ is invertible, it can be obtained that

$$\dot{y} = -(D_y g)^{-1}(D_x g)f \quad (16)$$

Then the equivalent system $\Sigma'$ is obtained as follows:

$$(\Sigma') \begin{cases} \dot{x} = f(x, y) \\ \dot{y} = -(D_y g)^{-1}(D_x g)f \end{cases} \quad (17)$$

The system $\Sigma'$ is well defined for $(x, y) \in R^{n+m}$, but except for $(x, y) \in \mathcal{S}$. So in order to analyze the system trajectories near the singular surface, Takens proposed a singular transformation. The idea of the technique is to multiply (17) with $\det(D_y g)$ to obtain a regularly transformed system $\Sigma''$ [24]:

$$(\Sigma'') \begin{cases} \dot{x} = f(x, y)\Delta(x, y) \\ \dot{y} = -\text{adj}(D_y g)(D_x g)f \end{cases} \quad (18)$$

where $\text{adj}(D_y g)$ denotes the adjoint matrix of $D_y g$. For the sake of simplicity, it may assume that $\det(D_y g) > 0$, then the system $\Sigma''$ is equivalent to $\Sigma'$ in $\mathcal{L} \backslash \mathcal{S}$. It is straightforward to observe that $\Sigma''$ and $\Sigma'$ differ in time scale and

$$\frac{dt}{d\tau} = \Delta(x, y) \quad (19)$$

where $\tau$ denotes the time scale of $\Sigma''$. Such time-scale transformation is also widely employed in singular perturbation methods [25]. The advantage of the transformed system $\Sigma''$ is that $\Sigma''$ is defined globally which allows us to use results of ODEs to analyze a DAE system.

To proceed, it is first necessary to define several important sets [6],[11]. The stability region of the stable equilibrium $(x_s, y_s)$ of the constrained system is denoted as $A(x_s, y_s)$, the boundary of the stability region is $\partial A$.

The pseudo-equilibrium set is defined as

$$\Psi = \{(x, y) \subset \mathcal{S} | \kappa(x, y) := \text{adj}(D_y g)(D_x g)f = 0\} \quad (20)$$

The semi-singular set is defined as

$$\Xi = \{(x, y) \subset (\mathcal{S} \backslash \Psi) | D_y \Delta \cdot \kappa = 0\} \quad (21)$$

Now we are ready to introduce the fundament result of the quasi-stability boundary of DAEs.

*Theorem 1* [12],[13]–For a constrained system as (12), let $W^S(x_i)$ denote the stable manifold of the anchor point $x_i$ on the boundary, then under mild conditions

$$\overline{\partial A(x_s)} = \overline{\bigcup_i W^S(z_i) \bigcup_j W^S(\Psi_j) \bigcup_l W^S(\Xi_l)} \bigcup (\mathcal{S} \cap \overline{\partial A(x_s)}) \quad (22)$$

where $z_i$ denotes the type-one hyperbolic equilibrium point on the stability boundary, $\psi_j$ denotes transverse pseudo-saddles on the stability boundary, $\Xi_l$ denotes the semi-saddles on the stability boundary.

The above result states that the stability boundary is composed of parts of the singular surface and the stable manifolds of certain anchor points (type-one hyperbolic equilibrium points, transverse pseudo-saddles and semi-saddles), refer to Fig. 9 in Section V.

#### B. Regularization Transformation Based on Minimum Modulus Eigenvalue

Evidently, the main assertion of *Theorem 1* is of geometrical nature (refer to Fig. 9 for a graphical illustration). In subsequent text, we develop some computational results in order to advance further research and engineering applications.

To begin with, notice that the magnitude of det($D_y g$) can be extremely high, this is undesirable from a numerical computation point of view. Now let us consider the expression of $D_y g$

$$D_y g = \begin{bmatrix} B & G \\ G & -B \end{bmatrix} - \begin{bmatrix} b & g \\ -g & b \end{bmatrix} = \begin{bmatrix} B-b & G-g \\ G+g & -B-b \end{bmatrix} \quad (23)$$

where

$$b = \mathrm{diag}\left[\frac{Q_i(V_{xi}^2 - V_{yi}^2) - 2P_i V_{xi} V_{yi}}{(V_{xi}^2 + V_{yi}^2)^2}\right]$$

$$g = \mathrm{diag}\left[\frac{P_i(V_{xi}^2 - V_{yi}^2) + 2Q_i V_{xi} V_{yi}}{(V_{xi}^2 + V_{yi}^2)^2}\right] \quad (24)$$

The matrix $B$ exhibits many properties of a Laplacian matrix [26], and it is weakly diagonally dominant and its order of magnitude is significantly larger than that of matrices $b$ and $g$. When matrix $g$ is disregarded, $D_y g$ is a real, symmetric matrix that is guaranteed to have real eigenvalues. Fig. 1 illustrates the distribution of eigenvalues of $D_y g$ for the system shown in section VI.

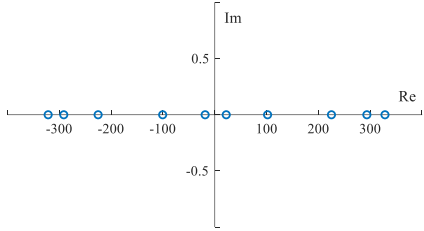

Fig. 1 Eigenvalues of $D_y g$

The figure illustrates that the eigenvalues of $D_y g$ are all real numbers and occur in pairs.

Taking Fig. 1 as an example, the eigenvalues shown above permit the calculation of the determinant of $D_y g$ to be approximately $-1.9 \times 10^{21}$. When $D_y g$ is close to singular, only one of the eigenvalues will be close to zero, while the product of the other eigenvalues will remain large. Consequently, when $D_y g$ is singular, our numerical example yields the determinant of $D_y g$ to be $-4.2 \times 10^{17}$, which is misleading. Besides, when calculating the eigenvalues of the Jacobian matrix of pseudo-saddles, MATLAB yields flawed non-zero eigenvalues which are superfluous, in addition to correct non-zero eigenvalues.

To resolve the above numerical instability, in what follows we propose a new transformation for $\Sigma'$ based on the minimum modulus eigenvalue of $D_y g$. To be precise, the minimum modulus eigenvalue mentioned above denotes the eigenvalue closest to zero. The new transformation employs the eigen-decomposition for $D_y g$

$$(D_y g)^{-1} = \frac{1}{\lambda_1} u_1 v_1^T + \cdots + \frac{1}{\lambda_n} u_n v_n^T \quad (25)$$

where $\lambda_1, \ldots, \lambda_n$ denote the eigenvalues of $D_y g$, $u_i$ and $v_i$ denote the eigenvectors of $D_y g$ with respect to $\lambda_i$. Assuming that the eigenvalue of $D_y g$ closest to 0 is $\lambda_1$ and $\lambda_1 > 0$ (if $\lambda_1 < 0$, $-\lambda_1$ can be used instead), then multiply (17) with $\lambda_1$ to obtain a regularly transformed system $\Sigma_\lambda$

$$(\Sigma_\lambda)\begin{cases} \dot{x} = f(x, y)\lambda_1 \\ \dot{y} = -\left(u_1 v_1^T + \frac{\lambda_1}{\lambda_2} u_2 v_2^T + \cdots + \frac{\lambda_1}{\lambda_n} u_n v_n^T\right)(D_x g)f \end{cases} \quad (26)$$

The system $\Sigma_\lambda$ has the same properties as $\Sigma''$, and the relationship of the time scale between $\Sigma_\lambda$ and $\Sigma'$ is

$$\frac{dt}{d\tau} = \lambda_1 \quad (27)$$

Compared with the transformed system $\Sigma''$, the new system $\Sigma_\lambda$ enjoys desirable numerical stability. In subsequent analysis, it will be shown (Section V) that, when utilizing the suggested system $\Sigma_\lambda$, one obtains simpler formulas.

Evidently, under the new transformation, *Theorem* 1 described in the previous sub-section is still applicable. And the new transformation helps to understand the pseudo-equilibrium set and the semi-singular set better. Notice that, in the spirit of equations (20) and (21), we have under the new transformation

$$\kappa(x, y) = \left(u_1 v_1^T + \frac{\lambda_1}{\lambda_2} u_2 v_2^T + \cdots + \frac{\lambda_1}{\lambda_n} u_n v_n^T\right)(D_x g)f \quad (28)$$

By definition every point in $\Psi$ is an equilibrium point of the transformed system $\Sigma_\lambda$, hence $\Psi$ is also the set of points satisfying the following equations

$$\begin{cases} g(x, y) = 0 \\ \lambda_1 = 0 \\ u_1 v_1^T (D_x g) f = 0 \end{cases} \quad (29)$$

It is easily understood that $\Psi$ is an $n - 2$ dimensional manifold.

The points in $\Xi$ are not pseudo-equilibrium points but whose vector field is tangent to $\mathcal{S}$. Therefore $\Xi$ is also the set of points satisfying the following equations

$$\begin{cases} g(x, y) = 0 \\ \lambda_1 = 0 \\ (D_y \Delta) u_1 v_1^T (D_x g) f = 0 \\ u_1 v_1^T (D_x g) f \neq 0 \end{cases} \quad (30)$$

Similarly, $\Xi$ is also an $n - 2$ dimensional manifold.

*C. System Trajectories for Angle Instability and Voltage Instability*

Apparently, a power system can exhibit distinct behavior in proximity to different parts of the stability boundary. Taking the familiar single-machine-infinity-bus system (see Fig. 2, and refer to Appendix A for detailed description of the model) as an example, Fig. 3 below illustrates the trajectory of the system whose post-fault state $x(t_{PF})$ is outside the stable manifold of controlling unstable equilibrium point (CUEP). The blue line denotes the system trajectories, and the red dotted line denotes the stability boundaries of the stable equilibrium point (SEP). It is seen from the figure that the post-fault system trajectory diverges from the stable equilibrium point and the simulation continues, indicating a rotor angle instability.

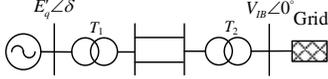

Fig. 2. A single-machine-infinity-bus system

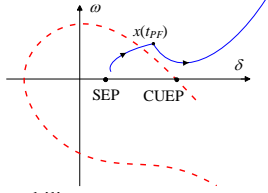

Fig. 3. Rotor angle instability

Fig. 4 shows the diagram of a single-machine system for voltage stability studies (Refer to Appendix B for detailed description of the model). Fig. 5 below shows the trajectory of the system which experiences short-term voltage instability. The blue line denotes the system trajectories, and the red dotted lines denote the stability boundaries which are composed of the manifold of pseudo-saddle and a piece of singular surface. It is seen that the system trajectory first passes the manifold of pseudo-saddle, which consists of part of the stability boundary, then moves on and terminates at the singular surface. This particular example shows that, to obtain a global picture of the short-term voltage behavior of a power system, one should ideally take into account the information of relevant anchor points, this is the focus of subsequent text.

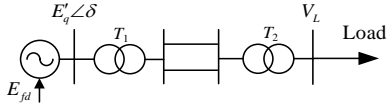

Fig. 4. A Single-machine system with a single load

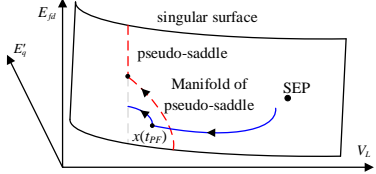

Fig. 5. Short-term voltage instability

The physical explanations for the two different instability phenomena mentioned above are as follows. In power systems, the algebraic equations represent the network power flow constraints. The singularity of the Jacobian matrix $D_y g$ indicates that the network equations lose solvability, meaning there is no feasible set of bus voltages that can satisfy the power injections under the given operating conditions. Physically, this corresponds to a situation where the power demand exceeds the system's ability to supply, often caused by excessive constant power loads or fault-induced system weakening. As the trajectory approaches the singular surface, voltage magnitudes at load buses start to collapse. Once the system reaches the singular surface, no voltage solution exists, leading to short-term voltage instability or collapse.

## IV. THE EXISTENCE AND COMPUTATION OF THE ANCHOR POINTS FOR POWER SYSTEMS DAE MODELS

It has been demonstrated that there are three types of anchor points on the stability boundary, namely type-one hyperbolic equilibrium points, pseudo-saddles and semi-saddles. The equilibrium points are closely related to rotor angle stability, and the stability margin can be quantified by utilizing the information of stability manifolds of them(refer to [18], [27], and references cited therein). Similarly, the pseudo-saddles and the semi-saddles are anchor points on the singular surface, and calculating their stability manifolds is an important way to quantify the stability margin of voltage stability. This section examines the existence of pseudo-saddles and semi-saddles on the stability boundary, and develops a method for computing the controlling pseudo-saddles.

### A. The Existence of Pseudo-Equilibrium Points

Based on (25), it has that
$$\text{adj}(D_y g) = \det(D_y g)(D_y g)^{-1}$$
$$= \lambda_1 \lambda_2 \cdots \lambda_n \left( \frac{1}{\lambda_1} u_1 v_1^T + \cdots + \frac{1}{\lambda_n} u_n v_n^T \right) \quad (31)$$
$$= \lambda_2 \cdots \lambda_n u_1 v_1^T + \ldots + \lambda_1 \lambda_2 \cdots \lambda_{n-1} u_n v_n^T$$

The result of (31) is still valid when $D_y g$ is not invertible. The reason is as follows.

When $D_y g$ is not invertible, assuming the zero eigenvalue of $D_y g$ is $\lambda_1$, on the basis of definition of adjoint matrix $A * \text{adj}(A) = \det(A) * I$, we have
$$(D_y g) * \text{adj}(D_y g)$$
$$= (\lambda_1 u_1 v_1^T + \cdots + \lambda_n u_n v_n^T)(\lambda_2 \cdots \lambda_n u_1 v_1^T + \ldots + \lambda_1 \lambda_2 \cdots \lambda_{n-1} u_n v_n^T)$$
$$= \lambda_2 \cdots \lambda_n (\lambda_2 u_2 v_2^T + \cdots + \lambda_n u_n v_n^T) u_1 v_1^T \quad (32)$$
$$= \lambda_2 \cdots \lambda_n (\lambda_2 u_2 v_2^T u_1 v_1^T + \cdots + \lambda_n u_n v_n^T u_1 v_1^T)$$
$$= O$$
$$= \det(D_y g) * I$$

Hence, the desired result follows.

According to the definition, at a pseudo-equilibrium point, one has $\lambda_1 = 0$, so
$$\text{adj}(D_y g) = \lambda_2 \cdots \lambda_n u_1 v_1^T \quad (33)$$

It follows that the set of pseudo-equilibrium points forms a high dimensional manifold satisfying the following equations:
$$v_1^T (D_x g) f = 0 \quad (34)$$

With a bit of abuse of notation, supposed there are $n$ generator buses and $m$ load buses in a power system, then it can be obtained that
$$D_x g = \begin{bmatrix} J_{IGx,\delta} & 0^{(n+m)\times n} & J_{IGx,E} & 0^{(n+m)\times n} \\ J_{IGy,\delta} & 0^{(n+m)\times n} & J_{IGy,E} & 0^{(n+m)\times n} \end{bmatrix} \quad (35)$$

where
$$J_{IGx,\delta} = \frac{\partial I_{Gx}}{\partial \delta} = -B_{LE}\text{diag}(E \circ \sin \delta)$$
$$J_{IGy,\delta} = \frac{\partial I_{Gy}}{\partial \delta} = -B_{LE}\text{diag}(E \circ \cos \delta)$$
$$J_{IGx,E} = \frac{\partial I_{Gx}}{\partial E} = B_{LE}\text{diag}(\cos \delta) \quad (36)$$
$$J_{IGy,E} = \frac{\partial I_{Gy}}{\partial E} = -B_{LE}\text{diag}(\sin \delta)$$

Then we get

$$(D_x g)f = \begin{bmatrix} \omega_0 J_{IGx,\delta}\omega + J_{IGx,E}\dot{E} \\ \omega_0 J_{IGy,\delta}\omega + J_{IGy,E}\dot{E} \end{bmatrix} \quad (37)$$

To show that pseudo-equilibrium point for power system DAE models always exists, one requires $(D_x g)f = 0$. On the singular surface, the changes in variables $\omega$, $E_{fd}$ and $E_{Cfd}$ have no effect on the singularity. This means that one can always find $\omega$ and $\dot{E}$ that satisfies $(D_x g)f = 0$, this is possible since

$$\dot{E} = \begin{bmatrix} \dot{E}'_q \\ \dot{E}_C \end{bmatrix} = \begin{bmatrix} T_{d0}'^{-1} \circ (E_{fd} - E'_q - (x_d - x'_d) \circ I_d) \\ K_i^{-1} \circ (E_{Cfd} + K_q \circ (Q_{ref} - Q_e)) \end{bmatrix} \quad (38)$$

This shows the existence of pseudo-equilibrium points for power system DAE model.

***Proposition 1*** – Consider the power system model by (1), (2) and (10), if $\mathcal{S} \neq \emptyset$, then $\Psi \neq \emptyset$.

Obviously, the set of pseudo-equilibrium points is a high dimensional manifold in state space, Section C will discuss how to compute a controlling pseudo-equilibrium point.

### B. Non-existence of Semi-Singular Points

According to the definition, the set of semi-singular points should satisfy the following equations:

$$(D_y \lambda_1)u_1 = \begin{bmatrix} v_1^T \dfrac{\partial D_y g}{\partial V_{x1}} u_1 & \cdots & v_1^T \dfrac{\partial D_y g}{\partial V_{ym}} u_1 \end{bmatrix} u_1 = 0 \quad (39)$$

To show that semi-singular points on voltage stability boundary may not exist, let us introduce

$$u_1 = \begin{bmatrix} u_{1x} \\ u_{1y} \end{bmatrix} \quad (40)$$

where $u_{1x}$ corresponds to the real part of network equation (10), and $u_{1y}$ corresponds to the imaginary part. Now expanding (39) yields

$$(D_y \lambda_1)u_1 = \sum_{i=1}\left(v_1^T \dfrac{\partial D_y g}{\partial V_{xi}} u_1\right)u_{1xi} + \sum_{i=1}\left(v_1^T \dfrac{\partial D_y g}{\partial V_{yi}} u_1\right)u_{1yi} = 0 \quad (41)$$

where

$$v_1^T \dfrac{\partial D_y g}{\partial V_{xi}} u_1 = -\begin{bmatrix} v_{1x}^T & v_{1y}^T \end{bmatrix} \begin{bmatrix} \dfrac{\partial b}{\partial V_{xi}} & \dfrac{\partial g}{\partial V_{xi}} \\ -\dfrac{\partial g}{\partial V_{xi}} & \dfrac{\partial b}{\partial V_{xi}} \end{bmatrix} \begin{bmatrix} u_{1x} \\ u_{1y} \end{bmatrix}$$

$$v_1^T \dfrac{\partial D_y g}{\partial V_{yi}} u_1 = -\begin{bmatrix} v_{1x}^T & v_{1y}^T \end{bmatrix} \begin{bmatrix} \dfrac{\partial b}{\partial V_{yi}} & \dfrac{\partial g}{\partial V_{yi}} \\ -\dfrac{\partial g}{\partial V_{yi}} & \dfrac{\partial b}{\partial V_{yi}} \end{bmatrix} \begin{bmatrix} u_{1x} \\ u_{1y} \end{bmatrix} \quad (42)$$

Apparently, the properties of eigenvector $u_1$ has significant impact on the value of $(D_y \lambda_1)u_1$. Notice that the matrix $D_y g$ has the following properties: 1. Since the transmission line resistance is much smaller than the line reactance, the elements of $G$ are much smaller than $B$; 2. $g$ is a sparse diagonal matrix whose elements are also much smaller than $B$; 3. Since $(B - b)$ is approximately a singular, irreducible M-matrix, there exists a vector $u > 0$ such that $(B - b)u = 0$ [28]. Thus the eigenvector $u_1$ of $D_y g$ has the following properties

$$u_{1x} \gg u_{1y}, u_{1x} > 0$$
$$v_{1x} \gg v_{1y}, v_{1x} > 0 \quad (43)$$

then (41) can be simplified to

$$(D_y \lambda_1)u_1 \approx -\sum_{i=1}\left(v_{1x}^T \dfrac{\partial b}{\partial V_{xi}} u_{1x}\right) u_{1xi} = -\sum_{i=1}\left(v_{1xi}^T \dfrac{\partial b_{ii}}{\partial V_{xi}} u_{1xi}\right) u_{1xi} \quad (44)$$

where

$$\dfrac{\partial b_{ii}}{\partial V_{xi}} = \dfrac{Q_i V_{xi} - P_i V_{yi}}{(V_{xi}^2 + V_{yi}^2)^2} - \dfrac{2V_{xi}[Q_i(V_{xi}^2 - V_{yi}^2) - 2P_i V_{xi} V_{yi}]}{(V_{xi}^2 + V_{yi}^2)^3} \quad (45)$$

By adjusting the voltage reference phase in the load area, it is possible to adjust the phase angle of the voltage close to 0, which means

$$V_{xi} \gg V_{yi} \quad (46)$$

Thus

$$\dfrac{\partial b_{ii}}{\partial V_{xi}} \approx \dfrac{-Q_i}{(V_{xi}^3)} < 0 \quad (47)$$

Then it can be concluded that

$$(D_y \lambda_1)u_1 \approx \sum_{i=1} -\left(v_{1xi}^T \dfrac{\partial b_{ii}}{\partial V_{xi}} u_{1xi}\right) u_{1xi} > 0 \quad (48)$$

It follows immediately that

$$(D_y \lambda_1)u_1 \neq 0 \quad (49)$$

This shows that semi-singular points on the voltage stability boundary may not exist.

***Proposition 2*** - Consider the power system model by (1), (2) and (10), if conditions (43) and (47) are met, then $\Xi = \emptyset$.

The result (49) is based on the assumption that $u_{1x} > 0$, and it also hold true when $u_{1x} < 0$.

To see an example, the elements of the right eigenvector matrix $u$ for test model named CSEE-VS (see [29] for detailed information) are calculated and the results are shown in Fig. 6.

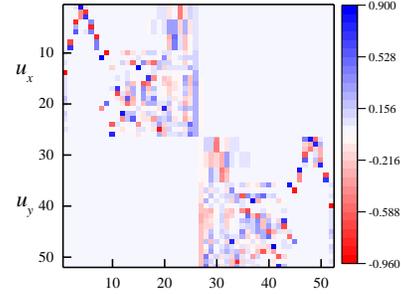

Fig. 6  The elements of the right eigenvector matrix

It can be seen from Fig. 6 that the right eigenvector matrix $u$ is a block diagonal matrix with nearly zero elements in its off-diagonal block which is consistent with the analysis in (43).

The result of $(D_y \lambda_1)u_1$ of the CSEE-VS test model is also calculated and is shown below in Fig. 7.

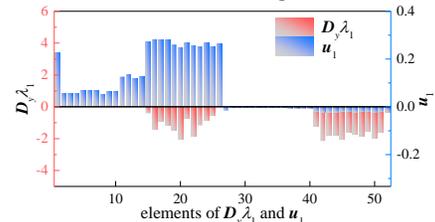

Fig. 7  The entries of $D_y \lambda_1$ and $u_1$ for CSEE-VS test system

It can be seen from Fig. 7 that $(D_y\lambda_1)u_1$ is strictly smaller than zero, that is, $(D_y\lambda_1)u_1 < 0$, this is consistent with the analysis presented above.

## C. Computation of The Controlling Pseudo-Saddle on The Stability Boundary

Controlling pseudo-saddle on the stability boundary plays a key role in determining the transient behavior of power systems. Theoretically, the controlling pseudo-saddle is defined dynamically as the first pseudo-saddle whose stable manifold is intersected by the fault-on trajectory, which is analogous to the definition of the CUEP [18]. However, in practice, since the pseudo-equilibrium set forms a high-dimensional surface containing infinitely many points, directly identifying this dynamically-defined controlling pseudo-saddle from simulation trajectories is numerically challenging. Therefore, a simplified numerical implementation is adopted in this paper as an alternative approach.

To proceed, let's assume that the system trajectory intersects a point $z_{sp}$ on the singular surface. The controlling pseudo-saddle $z_{cps}$ is then defined as the point in pseudo-saddle set that is closest to $z_{sp}$. In another word, it is the solution of the following optimization problem

$$\min_{z \in \Psi} \quad \|z - z_{sp}\|_2$$
$$s.t. \quad \begin{cases} g(x, y) = 0 \\ \lambda_1 = 0 \\ u_1 v_1^T (D_x g) f = 0 \end{cases} \quad (50)$$

where $\|z - z_{sp}\|_2$ denotes the 2-norm of vector $(z - z_{sp})$. The objective of the above model is to find $z$ which minimizes the Euclid distance function $\|z - z_{sp}\|_2$ and satisfies necessary constraints as Fig. 8 illustrates.

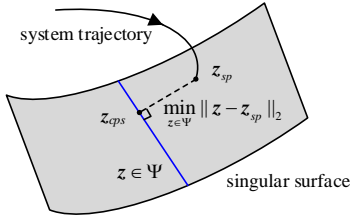

Fig. 8. Illustration of the optimization model for computing $z_{cps}$

## V. LOCAL REPRESENTATION OF THE MANIFOLD OF PSEUDO-SADDLES

This section describes a general method to calculate the local representation of the manifold of controlling pseudo-saddles. Application of the method to power system voltage stability is also detailed.

### A. Local Representation of Stable Manifold of Pseudo-Saddle

Similar to the stable manifold of type-one unstable equilibrium point, the stable manifold of $z_{cps}$ can be implicitly described as [30],[31]

$$W^s(z_{cps}) = \{z \mid d(z) = 0, f^T D_x d = \mu d, d(z_{cps}) = 0\} \quad (51)$$

where $d(z)$ denotes the stability manifold function, $\mu$ is the unstable eigenvalue of Jacobin matrix $J$ of the transformed system $\Sigma_\lambda$ at $z_{cps}$. Furthermore, the stable manifold of $z_{cps}$ has the following local approximate representation

$$\{z \mid d_P(z) = [z - z_{cps}]^T \eta = 0\} \quad (52)$$

where $d_p(z)$ denotes the local approximation of stability manifold function, $\eta$ denotes the left eigenvector which satisfies

$$\begin{cases} J^T \eta = \mu \eta \\ \eta^T \eta = 1 \end{cases} \quad (53)$$

It is worth noticing that, for a pseudo-saddle, the Jacobian matrix has at most two non-zero eigenvalues, one positive and one negative [12].

The local approximation of the stable manifold is derived based on first-order Taylor expansion of the system dynamics around the pseudo-saddle [32]. Therefore, the approximation error is of second order with respect to the distance from the pseudo-saddle. Specifically, the local error satisfies:

$$d(z) = [z - z_{cps}]^T \eta + O(\|z - z_{cps}\|^2) \quad (54)$$

This means the approximation is accurate when the system trajectory is close to the controlling pseudo-saddle $z_{cps}$, where linear terms dominate. As the trajectory moves further away, the error increases, making the estimation less accurate.

To find a local representation of the stable manifold of the pseudo-saddle, the key step is to compute the Jacobian matrix of the transformed system. To proceed, let $A$ be a matrix, $b$ and $c$ be vectors with appropriate dimensions, notice that

$$\frac{\partial (Ab)}{\partial c} = \left[ \frac{\partial A}{\partial c_1} b \cdots \frac{\partial A}{\partial c_n} b \right] + A \frac{\partial b}{\partial c} = \operatorname{col}(\frac{\partial A}{\partial c_i} b) + A \frac{\partial b}{\partial c} \quad (55)$$

where col(·) indicates that the elements within the parentheses are arranged in a columnar order. Now let $z = [x \; y]^T$, the Jacobian matrix $J$ of the transformed system $\Sigma_\lambda$ can be obtained as

$$J = \begin{bmatrix} \lambda_1 \dfrac{\partial f}{\partial z} + f \dfrac{\partial \lambda_1}{\partial z} \\ -\operatorname{col}(\Gamma_l D_x g f) - \sum_{i=1}^{n} \dfrac{\lambda_1}{\lambda_n} u_i v_i^T \dfrac{\partial D_x g f}{\partial z} \end{bmatrix} \quad (56)$$

where

$$\Gamma_l = \frac{\partial \lambda_1}{\partial z_l} \sum_{i=2} \frac{u_i v_i^T}{\lambda_i} - \sum_{i=2} \frac{\partial \lambda_i}{\partial z_l} \frac{\lambda_1 u_i v_i^T}{\lambda_i^2} + \sum_{i=1} \frac{\lambda_1}{\lambda_i} \left( \frac{\partial u_i}{\partial z_l} v_i^T + u_i \frac{\partial v_i^T}{\partial z_l} \right) \quad (57)$$

In particular, the Jacobian matrix $J$ of the transformed system $\Sigma_\lambda$ at a pseudo-saddle is

$$J = \begin{bmatrix} f \dfrac{\partial \lambda_1}{\partial z} \\ -\operatorname{col}(\Gamma_l D_x g f) - u_1 v_1^T \dfrac{\partial D_x g f}{\partial z} \end{bmatrix} \quad (58)$$

where

$$\Gamma_l = u_1 \frac{\partial v_1^T}{\partial z_l} + \frac{\partial \lambda_1}{\partial z_l} \sum_{i=2} \frac{u_i v_i^T}{\lambda_i} \quad (59)$$

The equation above requires the computation of eigenvalue derivatives and eigenvector derivatives. The derivative of the eigenvalue $\lambda_i$ of the matrix $J$ with respect to a parameter $p$ is

$$\frac{\partial \lambda_i}{\partial p} = v_i^T \frac{\partial J}{\partial p} u_i \tag{60}$$

Following the result in [33], the derivative of the right eigenvector $u_i$ of the matrix $J$ with respect to a parameter $p$ is

$$\frac{\partial u_i}{\partial p} = \sum_{j=1} \alpha_{ij} u_j \tag{61}$$

where

$$\alpha_{ik} = \begin{cases} -\frac{1}{\lambda_k - \lambda_i}\left(v_k^T \frac{\partial J}{\partial p} u_i\right) & k \neq i \\ -\sum_{j=1, \neq i} \alpha_{ij} u_j^T u_i & k = i \end{cases} \tag{62}$$

The derivative of the left eigenvector $v_i$ of the matrix $J$ with respect to a parameter $p$ is

$$\frac{\partial v_i}{\partial p} = \sum_{j=1} \beta_{ij} v_j \tag{63}$$

where

$$\beta_{ik} = -\alpha_{ki} \tag{64}$$

### B. Application to Power System Voltage Studies

The results presented in Section A are applicable to general DAEs. Application to power system voltage study is straightforward. In what follows the details are provided to help understand the structural properties of the Jacobian matrix and to provide convenience for future research. Since the derivative matrices of GFM converters are the same as those of synchronous machines, so the derivative matrices of the GFM converters are not repeated here.

*1) The expression of $\partial f / \partial z$*

The expression of $\partial f / \partial x$ is derived as

$$\frac{\partial f}{\partial x} = \begin{bmatrix} 0 & \omega_0 I & 0 & 0 \\ -J_{\omega\delta} & -\mathrm{diag}(M^{-1} \circ D) & -J_{\omega E} & 0 \\ -J_{E\delta} & 0 & -J_{EE} & \mathrm{diag}(T_{d0}'^{-1}) \\ 0 & 0 & 0 & -\mathrm{diag}(T_A^{-1}) \end{bmatrix} \tag{65}$$

where

$$\begin{aligned}
J_{\omega\delta} &= \mathrm{diag}[M^{-1} \circ x_d'^{-1} \circ E_q' \circ (V_{Gx} \circ \cos\delta + V_{Gy} \circ \sin\delta)] \\
J_{\omega E} &= \mathrm{diag}[M^{-1} \circ x_d'^{-1} \circ (V_{Gx} \circ \sin\delta - V_{Gy} \circ \cos\delta)] \\
J_{E\delta} &= \mathrm{diag}[T_{d0}'^{-1} \circ (x_d \circ x_d'^{-1} - I) \circ (V_{Gx} \circ \sin\delta - V_{Gy} \circ \cos\delta)] \\
J_{EE} &= \mathrm{diag}[T_{d0}'^{-1} \circ x_d \circ x_d'^{-1}]
\end{aligned} \tag{66}$$

The expression of $\partial f / \partial y$ is derived as

$$\frac{\partial f}{\partial y} = \begin{bmatrix} 0 & 0 & 0 & 0 \\ J_{\omega VGx} & 0 & J_{\omega VGy} & 0 \\ J_{EVGx} & 0 & J_{EVGy} & 0 \\ J_{EfdVGx} & 0 & J_{EfdVGy} & 0 \end{bmatrix} \tag{67}$$

where

$$\begin{aligned}
J_{\omega VGx} &= -\mathrm{diag}(M^{-1} \circ x_d'^{-1} \circ E_q' \cdot \sin\delta) \\
J_{\omega VGy} &= \mathrm{diag}(M^{-1} \circ x_d'^{-1} \circ E_q' \circ \cos\delta) \\
J_{EVGx} &= \mathrm{diag}[T_{d0}'^{-1} \circ (x_d - x_d') \circ x_d'^{-1} \circ \cos\delta] \\
J_{EVGy} &= \mathrm{diag}[T_{d0}'^{-1} \circ (x_d - x_d') \circ x_d'^{-1} \circ \sin\delta] \\
J_{EfdVGx} &= -\mathrm{diag}[T_A^{-1} \circ K_A \circ V_{Gx} \circ |V_G|^{-1}] \\
J_{EfdVGy} &= -\mathrm{diag}[T_A^{-1} \circ K_A \circ V_{Gy} \circ |V_G|^{-1}]
\end{aligned} \tag{68}$$

Then the expression of $\partial f / \partial z$ is obtained as

$$\frac{\partial f}{\partial z} = \begin{bmatrix} \frac{\partial f}{\partial x} & \frac{\partial f}{\partial y} \end{bmatrix} \tag{69}$$

*2) The expression of $\partial D_x g f / \partial z$*

The expression of $\partial D_x g f / \partial z$ is derived as

$$\frac{\partial D_x g f}{\partial z} = \mathrm{col}(\frac{\partial D_x g}{\partial z_i} f) + D_x g \frac{\partial f}{\partial z} \tag{70}$$

Based on (35), it is obvious that $D_x g$ is a function of rotor angles and the transient potentials, thus only the derivatives of $D_x g$ with respect to the rotor angles and the transient potentials are derived, and the expressions are as follows

$$\begin{aligned}
\frac{\partial D_x g}{\partial \delta_i} &= \begin{bmatrix} \mathcal{H}(-x_{di}'^{-1} E_{qi}' \cos\delta_i) & 0 & \mathcal{H}(-x_{di}'^{-1} \sin\delta_{ii}) & 0 \\ \mathcal{H}(x_{di}'^{-1} E_{qi}' \sin\delta_i) & 0 & \mathcal{H}(-x_{di}'^{-1} \cos\delta_i) & 0 \end{bmatrix} \\
\frac{\partial D_x g}{\partial E_{qi}'} &= \begin{bmatrix} \mathcal{H}(-x_{di}'^{-1} \sin\delta_i) & 0 \\ \mathcal{H}(-x_{di}'^{-1} \cos\delta_i) & 0 \end{bmatrix}
\end{aligned} \tag{71}$$

where $\mathcal{H}(\cdot)$ denotes a sparse matrix with elements only in the $i$-th row and the $i$-th column.

*3) The expression of $\partial D_y g / \partial z$*

Based on equation (23), the expression of $\partial D_y g / \partial z_i$ is

$$\frac{\partial D_y g}{\partial z_i} = -\begin{bmatrix} \frac{\partial b}{\partial z_i} & \frac{\partial g}{\partial z_i} \\ \frac{\partial g}{\partial z_i} & \frac{\partial b}{\partial z_i} \end{bmatrix} \tag{72}$$

As shown in (35), it can be seen that $D_y g$ is a function of bus voltages, thus only the derivatives of $D_y g$ with respect to the voltages are derived, and the expressions are as follow

$$\begin{aligned}
\frac{\partial b}{\partial V_{xi}} &= 2\mathcal{H}\{\frac{Q_i V_{xi} - P_i V_{yi}}{(V_{xi}^2 + V_{yi}^2)^2} - \frac{2V_{xi}[Q_i(V_{xi}^2 - V_{yi}^2) - 2P_i V_{xi} V_{yi}]}{(V_{xi}^2 + V_{yi}^2)^3}\} \\
\frac{\partial b}{\partial V_{yi}} &= 2\mathcal{H}\{\frac{-Q_i V_{yi} - P_i V_{xi}}{(V_{xi}^2 + V_{yi}^2)^2} - \frac{2V_{yi}[Q_i(V_{xi}^2 - V_{yi}^2) - 2P_i V_{xi} V_{yi}]}{(V_{xi}^2 + V_{yi}^2)^3}\} \\
\frac{\partial g}{\partial V_{xi}} &= 2\mathcal{H}\{\frac{P_i V_{xi} + Q_i V_{yi}}{(V_{xi}^2 + V_{yi}^2)^2} - \frac{2V_{xi}[P_i(V_{xi}^2 - V_{yi}^2) + 2Q_i V_{xi} V_{yi}]}{(V_{xi}^2 + V_{yi}^2)^3}\} \\
\frac{\partial g}{\partial V_{yi}} &= 2\mathcal{H}\{\frac{-P_i V_{yi} + Q_i V_{xi}}{(V_{xi}^2 + V_{yi}^2)^2} - \frac{2V_{yi}[P_i(V_{xi}^2 - V_{yi}^2) + 2Q_i V_{xi} V_{yi}]}{(V_{xi}^2 + V_{yi}^2)^3}\}
\end{aligned} \tag{73}$$

Having calculated the derivatives above, the complete local representation of the stability manifold of the pseudo-saddle can be obtained. Based on the local representation of the stability manifold of the pseudo-saddle, a numerical metric to quantify the voltage stability margin can be defined as follows

$$C_V = \frac{d_p(\boldsymbol{x}_{tf})}{d_p(\boldsymbol{x}_s)} \tag{74}$$

where $\boldsymbol{x}_s$ and $\boldsymbol{x}_{tf}$ denote the stable equilibrium point and the states after fault clearance, respectively. Apparently, if the metric $C_V$ is positive, the state at the end of the fault and the stable equilibrium point are on the same side of the stable manifold. Once the metric becomes negative, the system will move on and finally reach the singular surface.

The significance of metric (74) are two folds: first, to provide an approximate stability boundary; second, to reveal a new mechanism of voltage instability. The proposed mechanism is based on the pseudo-saddle located on the singular surface, which marks the loss of solvability of the algebraic network equations—a key feature of voltage instability. From (23) and (24), it can be seen that (23) is singular only when $\boldsymbol{b}$ and $\boldsymbol{g}$ increase. And the necessary condition for $\boldsymbol{b}$ and $\boldsymbol{g}$ to increase is a large drop in voltage, corresponding to a voltage collapse.

In contrast, angle instability is driven by the divergence of system trajectory around type-one hyperbolic equilibrium points, which are not involved in the pseudo-saddle dynamics. As shown in Fig. 3 and Fig. 5, once the trajectory crosses the pseudo-saddle manifold, it converges to the singular surface, leading to short-term voltage collapse rather than angle divergence.

Fig. 9 illustrates the stability boundary of a DAE system. As can be seen from Fig. 9, the point $z_{tf1}$ resides within the stability region, the stability estimation results provided by metrics $C_V$ and $\lambda_1$ are consistent. However, as the point $z_{tf2}$ resides outside the stability region, the metrics $C_V$ is able to predict the imminent loss of stability of the system with greater precision than $\lambda_1$.

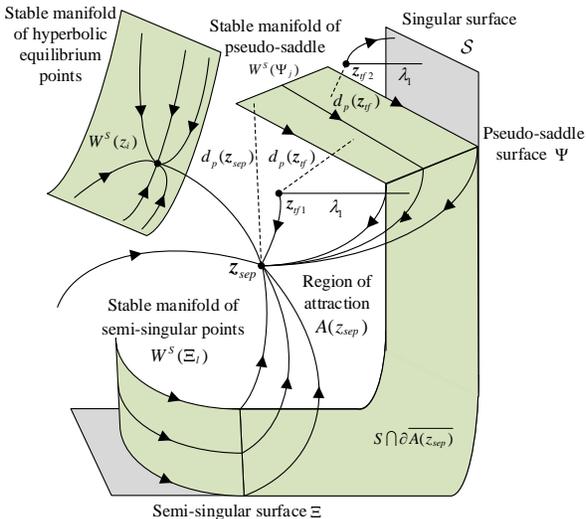

Fig. 9. The stability boundary of a DAE

## VI. SIMULATION RESULTS

This section presents case study results that serve to further validate the theoretical findings described in previous sections.

As shown in Fig. 10, the model of a power system with three synchronous machines and two loads (Taylor's test system) was established in the MATLAB/Simulink platform to perform simulation. The data of the power system is detailed in [34].

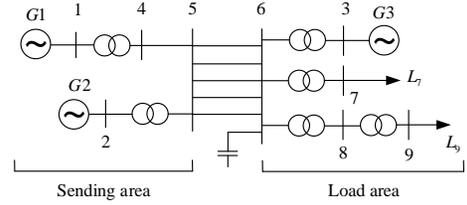

Fig. 10. A power system with three synchronous machines and two loads

### A. The Validity of The Proposed Transformation

This subsection verifies the validity of the transformed system $\Sigma_\lambda$ in both angle instability scenario (Fig. 11) and voltage instability scenario (Fig. 12).

In the power system mentioned above, the total system load is 6000 MW, of which the constant power loads share is 0%. At 0.2 s, a short circuit fault occurs at bus6. After 0.06 s, the fault is cleared by removing the faulted line between bus5 and bus6. After this fault, the system will experience angle instability. The trajectories of the DAE system and the transformed system $\Sigma_\lambda$ are presented in Fig. 11 below.

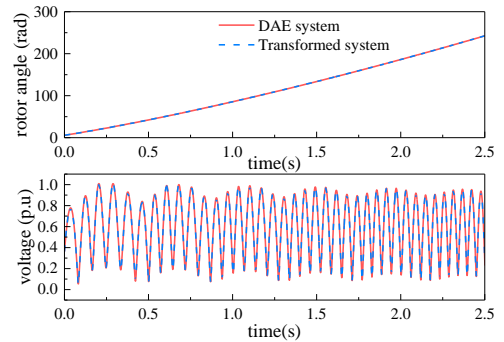

Fig. 11. Model validation when loss of synchronization occurs

Similarly, the total system load is 6000 MW, of which the constant power loads share now becomes 40%. At 0.2 s, a short circuit fault occurs at bus6. After 0.06 s, the line between bus5 and bus6 where the fault occurs is disconnected. After this fault, the system will experience voltage instability. The trajectories of the DAE system and the transformed system $\Sigma_\lambda$ are presented in Fig. 12 below.

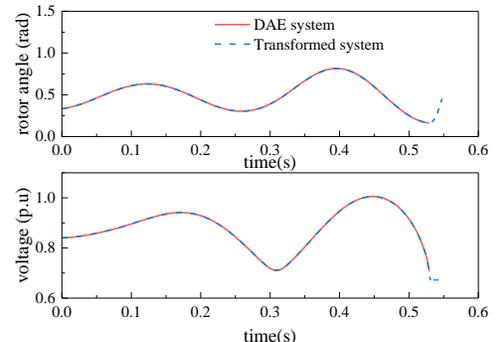

Fig. 12. Model validation when the system reaches the singular surface

The two figures above demonstrate that the proposed transformed model is applicable regardless of whether the system experiences angle instability or voltage instability.

## B. The Validity of The Proposed Voltage Stability Metric

In this subsection, the proposed voltage stability metric $C_V$ is compared with the minimum eigenvalues $\lambda_1$ to verify the accuracy of stability estimation. The total system load is now set to 6000 MW, of which the constant power loads share is 35%. At 0.2 s, a short circuit fault occurs at bus6. After 0.06 s, the fault is cleared by removing two lines between bus5 and bus6 where the fault occurs. The trajectories of the proposed metric $C_V$ and $\lambda_1$ described in our recent work [15] are presented in Fig. 13 below.

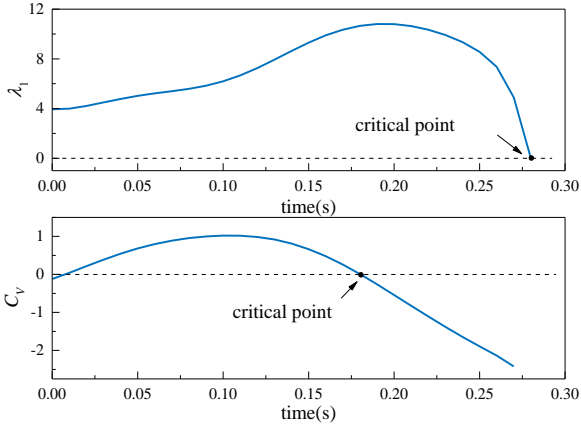

Fig. 13. Comparison between the metric $C_V$ and $\lambda_1$ presented in [15]

As can be seen from the figure, the proposed metric $C_V$ and minimum eigenvalue $\lambda_1$ behave roughly similarly. However, the trajectory of $C_V$ intersects the critical points at 0.175 s after the fault is cleared, while the trajectory of $\lambda_1$ intersects the critical point at 0.28 s after the fault is cleared, indicating that the stability estimation provided using $\lambda_1$ is overly optimistic. This is not to claim that $\lambda_1$ is not important, rather this example merely shows that the combined use of two metrics tends to yield improved stability estimation.

## C. The Non-Existence of the Semi-Singular Points for The Test System

In Section IV.B, the non-existence of semi-singular points was shown based on certain assumptions. In this subsection, the value of $\boldsymbol{D}_y\boldsymbol{g}$ is computed to verify the previous assertion and the result is shown below.

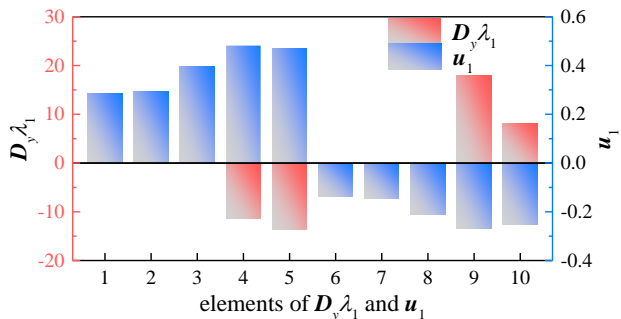

Fig. 14. The entries of $\boldsymbol{D}_y\lambda_1$ and $\boldsymbol{u}_1$ for Taylor's test system

From Fig. 14, it can be seen that the value of $\boldsymbol{D}_y\boldsymbol{g}$ is negative, indicating that the semi-singular points do not exist in the system shown in Fig. 10.

## D. The Impact of GFM Converters on Voltage Stability

In order to analyze the impact of GFM converters on the voltage stability of the power system, the system in Fig. 10 was adjusted by adding a converter-based generator at bus 10, and the modified system is shown in Fig. 15 (For detailed model, refer to Appendix C). The control diagram of VSM strategy is shown in Fig. 16 [35].

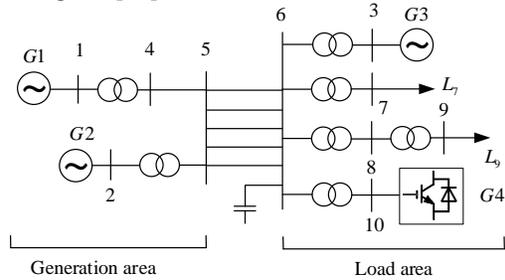

Fig. 15. A power system with a converter-based generator

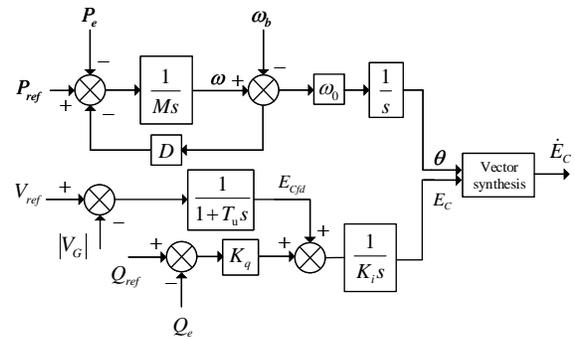

Fig. 16. Control block diagram of VSM strategy

The objective of the case study is to compare the voltage stability of the system when G4 is a GFL converter or a GFM converter, respectively. The total system load is still 6000 MW, of which the constant power loads share is 30%. At 0.2 s, a short circuit fault occurs at bus6. After 0.06 s, the fault is cleared by removing the faulted line between bus5 and bus6. In the event of such a fault, the system without converters connected is immediately unstable. For the sake of illustration, the active outputs of 200 MW, 400 MW and 600 MW for G4 are considered. In each case, the critical clearing time (CCT) was calculated and the results are shown in TABLE I.

TABLE I
CCT for different types of converter connected

| Type of converters | GFL converters | | |
|---|---|---|---|
| active output | 200MW | 400MW | 600MW |
| CCT/s | 0.01 | 0.05 | 0.12 |
| Type of converters | GFM converters | | |
| active output | 200MW | 400MW | 600MW |
| CCT/s | 0.40 | 0.42 | 0.43 |

From the results presented in TABLE I, it can be observed that the CCT of the system increases with the interconnection of renewable energy generators. Furthermore, the GFM converters have a much better impact on the CCT of the system.

## VII. CONCLUSIONS

In this work, some computational results for studying the voltage stability boundary of power systems interconnected with converters are presented. A new regularization transformation with desirable numerical stability is proposed.

Analytical properties of pseudo-saddles and semi-saddles, which are relevant to stability estimation in addition to singular surface, are examined. A local representation of the stable manifold of the pseudo-saddle is given. Based on the proposed voltage stability metric $C_V$, the impact of GFM/GFL converters on short-term voltage stability is studied. Future research may focus on, for example, the generalization of the presented results to systems with more detailed dynamics models.

The complex interaction between rotor angle instability and algebraic singularity warrants further study. Future work could establish a theoretical framework to clarify how severe rotor angle deviations drive systems toward the singular surface and explore conditions under which angle instability triggers or worsens voltage instability. Additionally, extending pseudo-saddle manifolds to consider interactions with type-one hyperbolic equilibrium points could bridge traditional transient stability analysis and singularity-based voltage stability assessment.

## APPENDIX

### A. The Model of the Single-Machine-Infinity-Bus System

The model of the single-machine-infinity-bus system is

$$f : \begin{cases} \dot{\delta} = \omega_0 \omega \\ M\dot{\omega} = P_m - \dfrac{E'_q V_{IB}}{x'}\sin\delta - D\omega \end{cases} \quad (A1)$$

where, $V_{IB}$ denotes the voltage of the infinity-bus, $x'$ denotes the sum of the line reactance, transformer reactance and $d$-axis transient reactance, and the interpretations of other symbols are the same as the model in Section II.

### B. The Model of The Single-Machine System with A Single Load

The model of the single-machine system with a single load system is

$$f : \begin{cases} \dot{\delta} = \omega_0 \omega \\ M\dot{\omega} = P_m - P_e - D\omega \\ T'_{d0}\dot{E}'_q = E_{fd} - E'_q - (x_d - x'_d)I_d \\ T_A \dot{E}_{fd} = K_A(V_{ref} - |V_G|) - E_{fd} \end{cases} \quad (B1)$$

$$g : \begin{cases} 0 = BV_x + GV_y + I_{Gx} - I_{Lx} \\ 0 = GV_x - BV_y + I_{Gy} - I_{Ly} \end{cases}$$

The interpretations of symbols are the same as the model in Section II.

### C. The Model of The System Shown in Fig. 15

The model shown in Fig. 15 is presented as follows:

For G1-G3, the models are presented by the model of synchronous machines:

$$f_1 : \begin{cases} \dot{\delta}_i = \omega_0 \omega_i \\ M_i \dot{\omega}_i = P_{mi} - P_{ei} - D_i \omega_i \\ T'_{d0i}\dot{E}'_{qi} = E_{fdi} - E'_{qi} - (x_{di} - x'_{di})I_{di} \\ T_{Ai}\dot{E}_{fdi} = K_{Ai}(V_{refi} - |V_{Gi}|) - E_{fdi} \end{cases} \quad (C1)$$

where $i$ = 1, 2, 3.

For G4, the model is presented by the model of GFM converter:

$$f_2 : \begin{cases} \dot{\delta}_4 = \omega_0 \omega_4 \\ M_4 \dot{\omega}_4 = P_{ref4} - P_{GFM4} - D_4 \omega_4 \\ K_{i4}\dot{E}_{C4} = E_{Cfd4} + K_{q4}(Q_{ref4} - Q_{e4}) \\ T_{u4}\dot{E}_{Cfd4} = V_{ref4} - |V_{G4}| - E_{Cfd4} \end{cases} \quad (C2)$$

The network equations are:

$$g : \begin{cases} 0 = BV_x + GV_y + I_{Gx} - I_{Lx} \\ 0 = GV_x - BV_y + I_{Gy} - I_{Ly} \end{cases} \quad (C3)$$

The interpretations of the symbols are the same as the model in Section II.

## Biographies

**Zhenyao Li** received the B.Eng. degree from the School of Electrical and electronic Engineering, Huazhong University of Science and Technology, Wuhan, China, in 2017, and the M.Eng. degree from the College of Electrical Engineering, Zhejiang University, Hangzhou, China, in 2021. He is currently working toward the Ph.D. degree with the College of Electrical Engineering, Zhejiang University, Hangzhou, China. His research interests include stability analysis and control of grid-connected power converters.

**Yifan Yao** received the B.Eng. degree in International Education Institute, North China Electric Power University, Baoding, China, in 2023. He is currently working toward the Ph.D. degree in the College of Electrical Engineering, Zhejiang University, Hangzhou. His research interests include power system stability analysis and control.

**Deqiang Gan** (SM'98) has been with the faculty of Zhejiang University since 2002. He visited the University of Hong Kong in 2004, 2005 and 2006. Deqiang worked for ISO New England, Inc. from 1998 to 2002. He held research positions in Ibaraki University, University of Central Florida, and Cornell University from 1994 to 1998. Deqiang received a Ph.D. in Electrical Engineering from Xian Jiaotong University, China, in 1994. He served as an editor for European Transactions on Electric Power (2007-2014). His research interests are power system stability and control.